# Do females create higher impact research? Scopus citations and Mendeley readers for articles from five countries[1]

Mike Thelwall, School of Mathematics and Computing, University of Wolverhampton, UK.

There are known gender imbalances in participation in scientific fields, from female dominance of nursing to male dominance of mathematics. It is not clear whether there is also a citation imbalance, with some claiming that male-authored research tends to be more cited. No previous study has assessed gender differences in the readers of academic research on a large scale, however. In response, this article assesses whether there are gender differences in the average citations and/or Mendeley readers of academic publications. Field normalised logged Scopus citations and Mendeley readers from mid-2018 for articles published in 2014 were investigated for articles with first authors from India, Spain, Turkey, the UK and the USA in up to 251 fields with at least 50 male and female authors. Although female-authored research is less cited in Turkey (-4.0%) and India (-3.6%), it is marginally more cited in Spain (0.4%), the UK (0.4%), and the USA (0.2%). Female-authored research has fewer Mendeley readers in India (-1.1%) but more in Spain (1.4%), Turkey (1.1%), the UK (2.7%) and the USA (3.0%). Thus, whilst there may be little practical gender difference in citation impact in countries with mature science systems, the higher female readership impact suggests a wider audience for female-authored research. The results show that the conclusions from a gender analysis depend on the field normalisation method. A theoretically informed decision must therefore be made about which normalisation to use. The results also suggest that arithmetic mean-based field normalisation is favourable to males.

## 1 Introduction

Academia has a long history of gender imbalances. Female undergraduates in the USA are still a minority in Science, Technology Engineering and Maths (STEM) and a majority of health care, elementary education and the domestic sphere (HEED) subjects (Tellhed, Bäckström, & Björklund, 2017). Gender differences vary internationally. For example, computing is male dominated in the USA but attractive for females in India (Gupta, 2012) and there is approximate computing gender parity in Malaysia from undergraduate level to department heads (Othman & Latih, 2006). Several studies point to females publishing fewer articles (Aksnes, Rorstad, Piro, & Sivertsen, 2011; Ceci, Ginther, Kahn, & Williams, 2014; Nielsen, 2016; Rørstad & Aksnes, 2015) but females are more frequently in jobs with a higher teaching requirement so if this is not considered then the difference will be exaggerated (Ceci, Ginther, Kahn, & Williams, 2014; Xie & Shauman, 1998). For example, for six ecology journals 1994-2008, removing career breaks and self-citations eliminated the productivity advantage of male researchers (Cameron, White, & Gray, 2016). Whilst academic gender imbalances were initially due to banning women, then to prejudice against females, at least in some countries and fields, they may now be primarily the result of gender differences in life choices and social pressures (Ceci & Williams, 2011). Nevertheless, some scientific fields may be disadvantaged by a shortage of females, either because of a consequent lowering of standards (due to less competition) or from missing female perspectives or female-associated capabilities. In countries where prejudice is not the cause of gender imbalances then hypothesised explanations include gendered topic preferences (Su & Rounds, 2015) and field-specific gendered cultures (Cheryan, Ziegler, Montoya, & Jiang, 2017). It is important to fully understand gender differences in research because it is easy for well-meaning actions to exacerbate rather than ameliorate the problem (Caffrey, Wyatt, Fudge, Mattingley, Williamson, & McKevitt, 2016). This is also important in the context that the use of standard bibliometric indicators having the potential to favour males in allocating funding, exacerbating gender differences (Abramo, Cicero, & D'Angelo, 2015; Nielsen, 2017).

---

[1] *Journal of Informetrics*, in press

Research impact is an important dimension of gender differences because if female-authored research is undervalued then this would lead to female researcher career attrition and fewer opportunities for promotion. This might occur due to prejudice against female-authored research by citers or if career breaks and part-time working were associated with lower impact research (Brooks, Fenton, & Walker, 2014), disproportionately affecting females. A previous large-scale study of differences between field-normalised citation rates for female-authored and male authored journal articles 2011 to 2015 (including articles with any author of each gender) found them to be similar overall in the 27 fields examined (Elsevier, 2017). Female-authored articles received marginally more citations in the USA and UK but fewer in Australia, Canada, and the EU28 (Elsevier, 2017, p. 31). Spain, India and Turkey were not included (see below). This study used arithmetic mean-based calculations, which are inappropriate for highly skewed citation data. It also used relatively broad categories, and so may be biased against one gender if it is overrepresented in low citation specialisms within broad fields. For the same article sets, field-normalised download rates were higher for female-authored research in most countries examined, including the UK and USA (Elsevier, 2017, p. 31). This raises the possibility that female-authored research tends to have more non-academic impact than male-authored research. In contrast, another study found female first-authored research to be less cited overall (Larivière, Ni, Gingras, Cronin, & Sugimoto, 2013; also for 12 disciplines using aggregated data from 2008 to 2015: Larivière & Sugimoto, 2017) and published in journals with lower Journal Impact Factors (JIFs) (Larivière, & Sugimoto, 2017). This study also used arithmetic means, relatively broad categories for normalisation, some of the data is a decade old (from 2008) and countries were reported as a single group.

The relationship between author gender and research impact has been investigated in a range of countries and fields. In four Norwegian universities, female-authored Web of Science (WoS) articles and reviews 2005-2008 were slightly (7%) less cited. Broken down by role, female professors were 10% less cited and postdocs were 16% less cited, but there was little difference for associate professors and PhD students (Aksnes, Rorstad, Piro, & Sivertsen, 2011). Males were substantially more cited in Humanities (49%, n=253) and Mathematics (23%, n=225), but less cited in Computer Science/Information Science (-16%, n=296). This study normalised at the level of fields (also by journal, when the difference reduced to 3%) and separately for articles and reviews. It used a Norwegian register of researchers to determine gender and allocated an article to a gender if at least one of its authors had that gender. Other studies have had conflicting results. Male first-authored papers in five astronomy journals 1996-2015 were 6% more cited (Caplar, Tacchella, & Birrer, 2017) but there was no difference for nanotechnology research 2005-7 (Sotudeh & Khoshian, 2014). There was a male advantage in field normalised citation impact for staff publications from 2009 at one Danish university, except for the life sciences (Nielsen, 2016). An investigation of factors that might influence citation rates of Quebec researchers found Natural Sciences and Engineering to be gender neutral but largely female teams tended to be less cited in Health Sciences (Beaudry & Larivière, 2016). For Italian research assessment, bibliometric indicators favour males less than peer review (Jappelli, Nappi, & Torrini, 2017).

One reason why male-authored research may be more cited is that male authors tend to self-cite substantially more. For JSTOR papers 1990-2011, the male self-citation rate was 70% higher and consistently higher across all broad fields investigated (King, Bergstrom, Correll, Jacquet, & West, 2017). This study inferred author gender from first names using historic US census data. It is not clear whether higher male self-citation represents narrower research careers for males (hence a greater likelihood of relevant prior work to cite), self-promotion through self-citation or another cause. No gender self-citation difference was found in a study of astronomy journals (Caplar, Tacchella, & Birrer, 2017). Another reason why males might be cited more is that researchers that author more papers over their career tend to produce higher impact work, disproportionately advantaging males for higher productivity, dominance of senior roles, and fewer career breaks. A study of WoS publications 2008-11 of Swedish researchers using first name heuristics to classify gender and fractional authorship counting found that more productive researchers of both genders (separately) tended to have a higher proportion of more highly cited papers (van den Besselaar, & Sandström, 2017). This could be partly an attrition effect, if researchers are motivated to publish more when their previous research has had an impact.

Despite the above research into gender differences in citation counts, almost nothing is known about gender differences in readership. Mendeley (Gunn, 2014) is the best source of readership information because it is used by 12-20% of researchers (Van Noorden, 2014) and is more popular than alternative social reference sharing sites. It is more common than other altmetrics, except perhaps Twitter, and has higher correlations with citation counts (Costas, Zahedi, & Wouters, 2015; Thelwall, Haustein, Larivière, & Sugimoto, 2013). Download data, which would be an alternative type of readership evidence (although excluding print journal readers), is not available for all articles and only reports total downloads and not information about the downloaders. Most people register articles in Mendeley because they have read them or intend to read them (Mohammadi, Thelwall, & Kousha, 2016) so it gives information about readership. There are high long-term correlations between Mendeley readership counts and citation counts in almost all fields (Thelwall, 2017b) and most readers are students, researchers or lecturers (Mohammadi, Thelwall, Haustein, & Larivière, 2015) and so Mendeley readership counts predominantly reflect citation-like interest. There is an element of national self-interest in Mendeley readership, with article readers disproportionately originating from the same countries as the article authors (Thelwall & Maflahi, 2015). An investigation of all WoS publications of 494 European and Israeli astronomers and astrophysicists found slightly more readers per article for the publications of the males in this list (Bar-Ilan & van der Weijden, 2014). This study did not normalise for publication year and researcher age. An analysis of WoS papers from 2012 using 134 field-based National Science Foundation journal categories found that news, blog and tweet altmetrics (not including Mendeley) did not give a clear male or female attention advantage overall (Paul-Hus, Sugimoto, Haustein, & Larivière, 2015).

This article assesses whether gender differences in citation impact persist when using Scopus data (most of the above studies used WoS), when aggregated at the narrow field level (the above studies used broad fields except for some using 134 narrow fields), for multiple countries analysed separately, when using appropriate averaging, and for more recent data. It also assesses gender differences in Mendeley readers on the same basis. The following questions drive the study. Although the first question has previously been addressed, this article reassesses it with different methods (narrow fields, newer data, skew-sensitive averaging). The final question addresses a secondary issue. Since Mendeley readers can be used for early impact evidence, it is useful to assess whether this would give reasonable answers if the factor of interest is author gender.

- RQ1: Is there a difference in the number of citations attracted by articles authored by males and females? Does this vary by country?
- RQ2: Is there a difference in the number of Mendeley readers attracted by articles authored by males and females? Does this vary by country?
- RQ3: Can Mendeley readers be used as a proxy for Scopus citations when investigating gender differences?
- RQ4: Are citation impact gender differences affected by the field normalisation method used?

## 2 Methods

The research design was to obtain sets of articles with known first author gender from all narrow fields and five countries from 2014 and compare average Scopus citation counts and Mendeley reader counts between male first-authored and female first-authored articles in each set. First author gender was used because the first author performs most of the work in most or all areas of scholarship (Larivière, Desrochers, Macaluso, Mongeon, Paul-Hus, & Sugimoto, 2016), with partial exceptions within specialisms sometimes using alphabetical order: mathematics and economics (Levitt & Thelwall, 2013). Scopus was used as a more comprehensive source than the Web of Science (Mongeon & Paul-Hus, 2016), which was important for India, although both Web of Science and Scopus have relatively low coverage of non-English and arts and humanities journals (Mongeon & Paul-Hus, 2016). Articles from 2014 were investigated to give recent data that is nevertheless old enough to allow three years of citations. Whilst two years has previously been argued to be sufficient (Abramo, Cicero, & D'Angelo, 2011), there is no consensus and three years is both safer (given that some fields analysed will have slow citation growth rates) and more standard. Correlations are more difficult to interpret for sets of articles that are too recent, a side-effect of discrete data (Thelwall, 2016). A recent year is important because gender participation rates change over time and so old data would give misleading evidence about current practice. Narrow fields were analysed because mixing fields together would increase the chance that any positive results were due to differing audiences for the more male or female specialist topics within the set analysed.

The USA and UK were chosen to investigate as two major science-producing countries with similar research cultures. The USA is probably the most investigated country for gender and its similarity with the UK (shared language, partly shared culture, similar levels of economic development) can help assessing whether the USA results are nation-specific. Spain was selected as a non-English speaking country with a major science base and Turkey was selected as another non-English speaking country with a growing science base but with an unusual gender profile. Although a lower proportion of females work compared to the European average, it has an above average share (28% vs. 15% for Europe) of female full professors (Ucal, O'Neil, & Toktas, 2015), including for STEM subjects (Inanc & Özcan, 2016). Finally, India has a large and rapidly-growing economy, but with high gender inequality in education, according to the UNDP (2016), and culture-specific concerns about the position of women in science (e.g., Gupta, 2017). Only large countries were selected to ensure sufficient data to analyse.

## 2.1 Data

Scopus articles with an author from India, Spain, Turkey, the UK and the USA in 2014 were downloaded in May/June 2018 with queries of the following form, using the Scopus API.

SUBJMAIN(3402) AND DOCTYPE(ar) AND SRCTYPE(j) AND AFFILCOUNTRY("Turkey")

This query was submitted separately for each of the 310 distinct Scopus narrow fields and for each of the five countries. For example, 3402 is the code for the narrow field *Equine* within the broad field *Veterinary*. Articles for which the first author was not from the designated country were discarded. Mendeley readers were obtained for each article from the Mendeley API in Webometric Analyst in May and June 2018, combining a DOI search with a title/author/year search for optimal results (Zahedi, Haustein, & Bowman, 2014). Articles not found in Mendeley were assumed to have 0 readers.

The gender of article first authors was guessed from their first name, when present, using lists of common male and female first names. For the USA and UK the lists were taken from the USA census 1991, including male and female names from the 10,000 most popular names with at least 90% used by one gender. For the other four countries, Gender-API.com was used to detect first author genders, using their affiliation country as a parameter, and only accepting name judgments when the API reported as least 50 examples of the name analysed and at least 90% to be the same gender. This ensures that the results will be at least 90% accurate for researchers that have a gender inferred. Nevertheless, this conservative approach leaves most authors without an inferred gender. Typical reasons for a gender being unassigned include gender-neutral full or short form names (e.g., Pat, Sam, Jatinder, Noor, Álex), rare names (e.g., Mihee in the USA), and names from cultures that rarely use gendered names (e.g., Sikh). Articles with unknown gender first authors were discarded. The main side-effect of this procedure is that the results are likely to largely ignore the contributions of minority cultures (including international researchers) in each country.

Fields with less than 50 male or less than 50 female authors from any given country were discarded for that country (except for the average calculations) to reduce the chance of outliers in the results due to low numbers of papers from either gender.

## 2.2 Analysis

For each narrow field, the geometric mean number of readers was calculated for the male and female authored articles. Geometric means were used because citation counts (de Solla Price, 1976; Seglen, 1992) and readership counts (Thelwall & Wilson, 2016) are highly skewed and so the arithmetic mean is unreliable (Thelwall & Fairclough, 2015; Zitt, 2012). Since geometric means use a log transformation that cannot be applied to zeros, 1 was added to all readership counts and citation counts at the start and subtracted from the geometric mean at the end, a standard approach (Thelwall, 2017a).

Overall gender differences were calculated for each country using the Mean Normalised Log Citation Score (MNLCS) (Thelwall, 2017c) variant of the standard MNCS field normalised indicator (Waltman, van Eck, van Leeuwen, Visser, & van Raan, 2011ab). This calculation uses the natural log of citation counts to reduce skewing. It normalises each logged citation (or reader) count by dividing by the field average logged reader (or citation) count before averaging all these normalised log citation counts. This indicator is not unduly affected by highly skewed data and normalises for field differences in citation rates and so it is a relatively fair way to average the reader or citation counts of large sets of articles. MNLCS values were calculated separately for each country so that a value of 1 indicates average impact for that country rather than the world. This approach avoids practices in other countries obscuring each nation's results. The number of fields with higher male or female MNLCS values was also calculated for additional context. For this, only fields with at least 50 male first-authored and at least 50 female first-authored articles were used to reduce the influence of small fields.

The Mendeley and Scopus geometric means for each gender were then correlated between and across genders within each country to investigate the relationship between them. Pearson correlations were used because the sets of geometric means were not highly skewed (only the USA Mendeley readers of male first-authored articles were skewed above 3, at 3.3).

## 3 Results

For all countries, there were at least 82 fields with a minimum of 50 male first-authored and 50 female first-authored articles, giving a substantial range of scientific fields to compare across (Table 1). Male first-authored articles dominate all countries, but this may be an artefact of the first name gender detection procedure, which may be more effective for males if females tend to have rarer names. The gender detection procedure leaves varying proportions of articles ungendered. It is particularly ineffective in the UK, which may be due to 30% of academics being non-UK nationals (HESA, 2018, Figure 5) and therefore having names that may not conform to UK cultural norms.

Table 1. Summary statistics for the gendered Scopus journal articles from 2014 in the data set.

| Data set | India | Spain | Turkey | UK | USA |
|---|---|---|---|---|---|
| Fields with at least one gendered article | 275 | 296 | 296 | 300 | 302 |
| Fields with at least 50 male and 50 female first-authored articles | 99 | 158 | 82 | 177 | 251 |
| Female first-authored articles (as a percentage of all gendered articles) | 19165 (27%) | 26301 (42%) | 12749 (35%) | 33766 (38%) | 141407 (38%) |
| Male first-authored articles (as a percentage of all gendered articles) | 52152 (73%) | 36854 (58%) | 23391 (65%) | 55920 (62%) | 229620 (62%) |
| Ungendered articles (as a percentage of all articles) | 82698 (54%) | 57551 (48%) | 17981 (33%) | 137484 (61%) | 389300 (51%) |
| Female first-authored articles in fields with at least 50 male and 50 female first-authored articles (as a percentage of all gendered articles in these fields) | 15984 (27%) | 23706 (43%) | 9338 (37%) | 31059 (39%) | 138904 (38%) |
| Male first-authored articles in fields with at least 50 male and 50 female first-authored articles (as a percentage of all gendered articles in these fields) | 42350 (73%) | 31200 (57%) | 15914 (63%) | 49350 (61%) | 225276 (62%) |

## 3.1 Citation impact

For three countries there is a tiny female overall citation advantage (Figure 1): Spain (0.4%); UK (0.4%), and USA (0.2%). Both Turkey (-4.0) and India (-3.6) have male citation advantages, and these are more substantial. At the level of individual narrow fields, the results broadly echo the MNLCS overall results, with similar numbers of fields having higher citation rates for male first-authored and female first-authored articles, but many more fields having a male citation advantage in both India and Turkey (Figure 2).

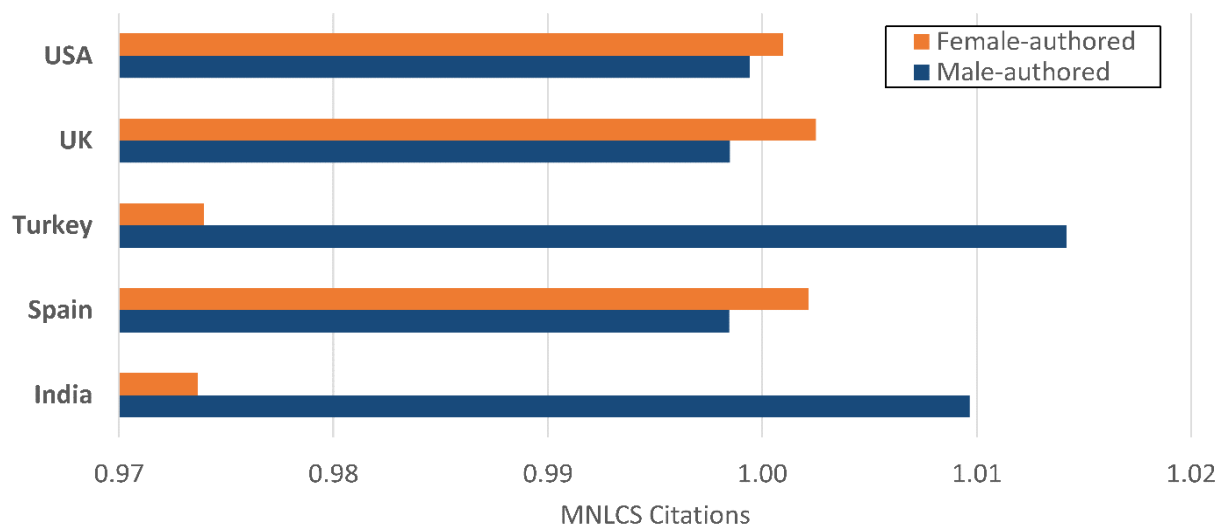

Figure 1. Scopus citations MNLCS for male first-authored and female first-authored articles calculated separately for each country, incorporating up to 301 narrow fields.

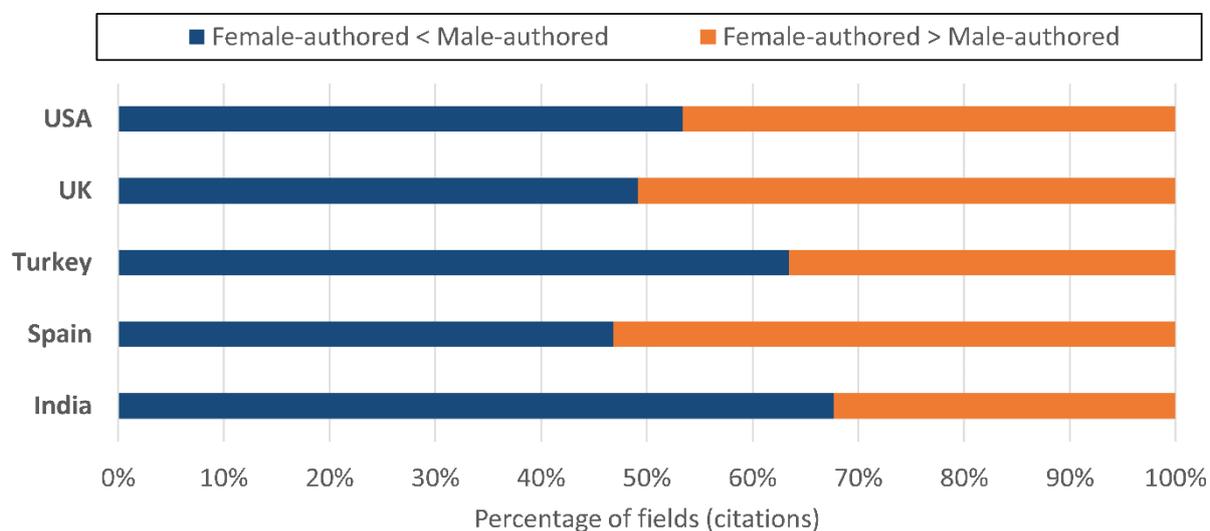

Figure 2. The percentage of fields in each country with a higher Scopus citations MNLCS for males or females (qualification: at least 50 male first-authored and at least 50 female first-authored articles in each field).

If the results are repeated but the data is normalised by broad field (n=26) rather than narrow field, then the citation advantages have the same direction but are wider for all countries except India: Spain (1.0%); UK (0.7%); USA (0.4%); Turkey (-4.3%); India (-1.6%).

Expressed differently, except in the case of India, the female citation advantage is greater when broad field aggregation is used.

If the results use MNCS instead of MNLCS then the results shift substantially away from females, with a male citation advantage in all countries (Figure 3): India (-7.2%); Spain (-2.3%); Turkey (-12.5%); UK (-5.9%), and USA (-6.0%).

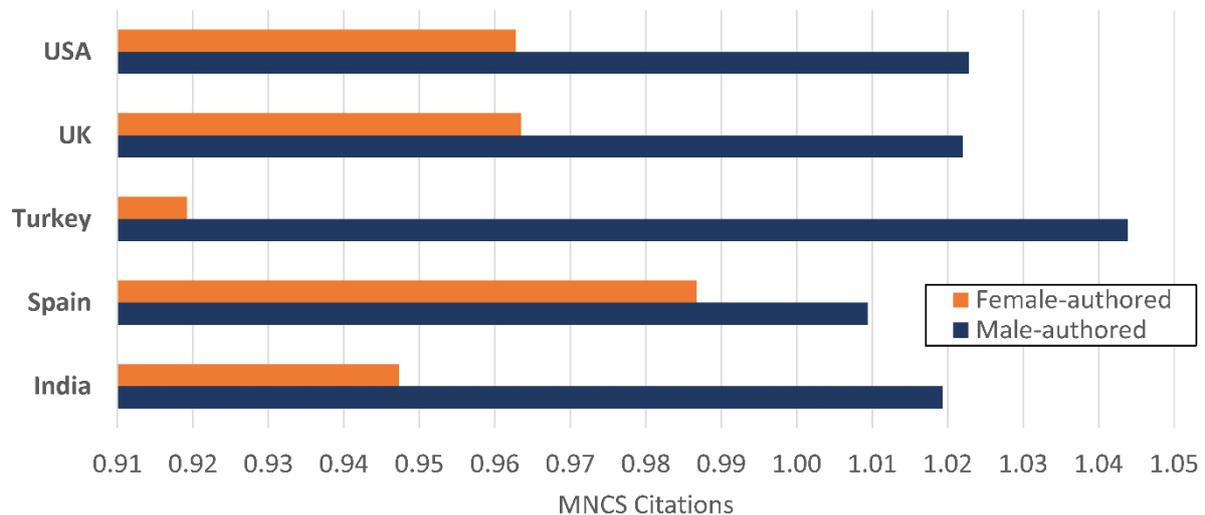

Figure 3. As Figure 1 for MNCS.

## 3.2 Mendeley reader impact

For Mendeley, all countries except India (-1.1%) have a female Mendeley readership advantage in the order of a few percentage points (Figure 4): Spain (1.4%); Turkey (1.1%), UK (2.7%), USA (3.0%). For all countries, the relative impact of female-authored research to male-authored research is higher for Mendeley readers than for Scopus citations, showing that there is a female shift in Mendeley relative to Scopus, whatever the cause. The per field results (Figure 5) echo the overall results.

If the results are repeated but the data is normalised by broad field (n=26) rather than narrow field, then the reader advantages have the same direction but are wider for all countries except India: Spain (1.7%); Turkey (1.7%); UK (4.1%); USA (4.1%); India (-0.4%). In all countries, using broad fields rather than narrow fields therefore increases the relative readership impact of female first-authored research.

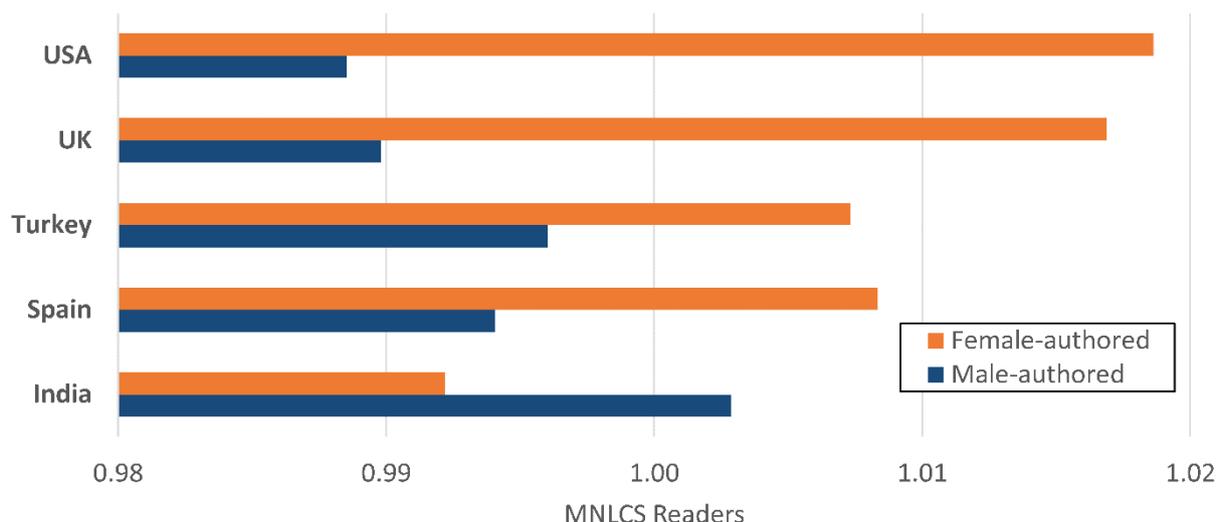

Figure 4. Mendeley readers MNLCS for male first-authored and female first-authored articles calculated separately for each country, incorporating up to 301 narrow fields.

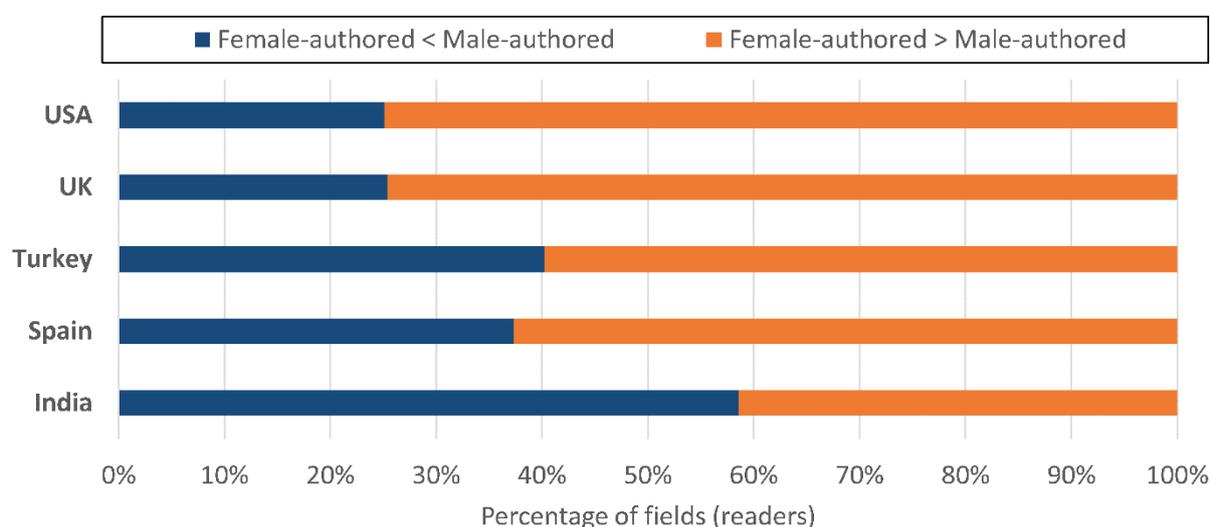

Figure 5. The percentage of fields in each country with a higher Mendeley readers MNLCS for males or females (qualification: at least 50 male first-authored and at least 50 female first-authored articles in each field).

### 3.3 Relationship between Mendeley readers and Scopus citations

There is a very high correlation (about 0.9) between the geometric means for male and female articles in each field, whether considering Scopus citations or Mendeley readers, and almost irrespective of country (exception: a correlation of 0.75 for Mendeley in Turkey) (Figure 6). When citations are compared to readers, the correlations are less strong, but still high (0.45-0.7). Comparing the male-female percentage difference (top set of bars in Figure 2), the Mendeley difference correlates highly (0.6-0.7) with the Scopus difference, showing that the two sources tend to give similar relative magnitude gender difference estimates.

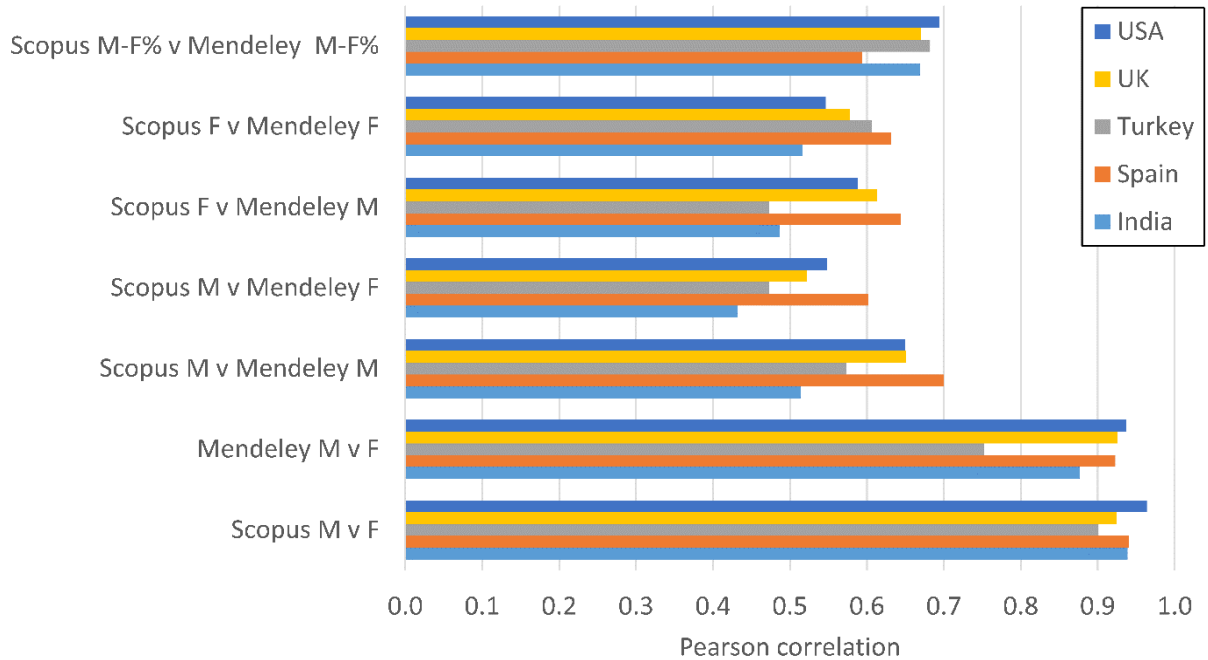

Figure 6. Pearson correlations between geometric mean Scopus citation and/or Mendeley reader counts for each narrow field (at least 50 male and 50 female articles in a field, giving n=82 to 251 narrow fields: see Table 1).

The country with the lowest correlation between Mendeley gender difference and Scopus gender difference, Spain, still has a strong correlation with few substantial outliers (Figure 7). For example, male-authored Spanish Virology research receives about 36% more citations and readers than female-authored Spanish Virology research. The biggest Spanish outlier is Hepatology (167 articles), with Male-authored research receiving 40% fewer citations but 4% more readers. Thus, in this case there is little gender difference from the perspective of Mendeley but male-authored Hepatology research has substantially less impact in Spain (i.e., the opposite direction to the normal bias).

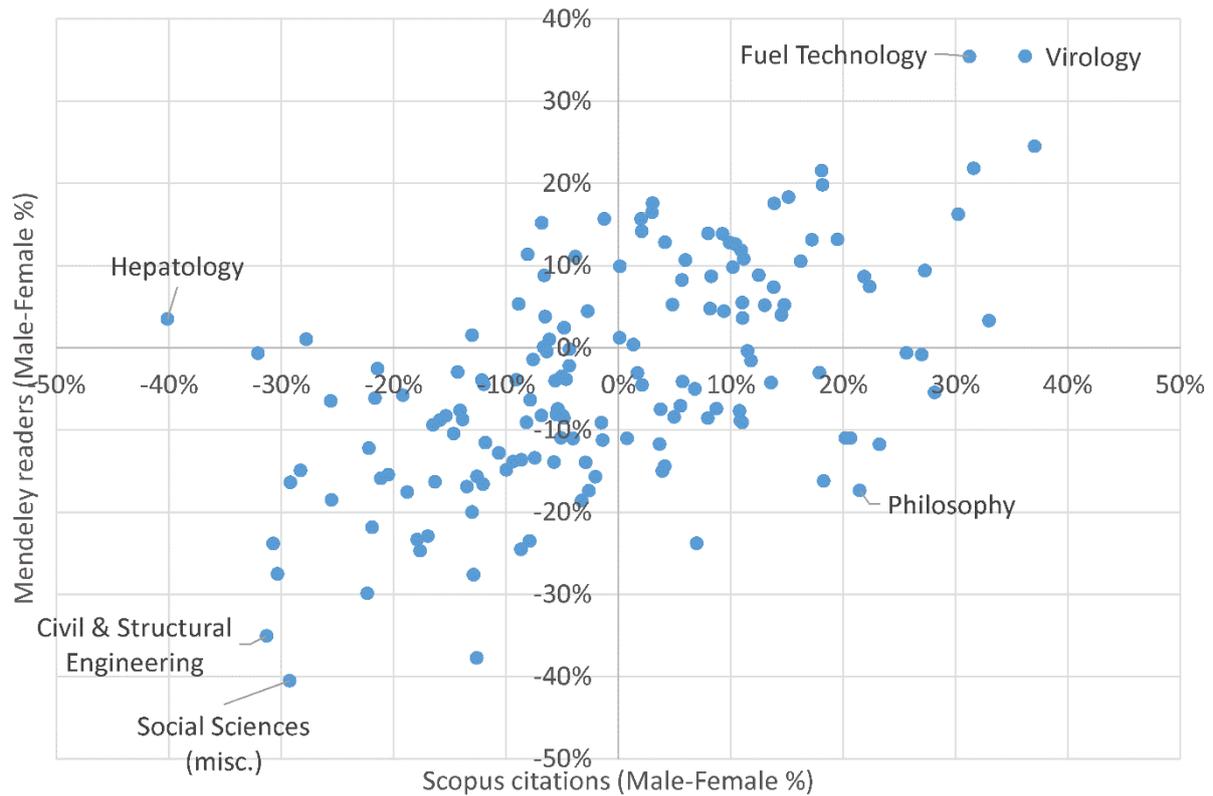

Figure 7. The average number of Mendeley readers per article difference between male-first-authored and female-first-authored Spanish articles from 2014 by narrow field (qualification: at least 50 male and 50 female articles in a field; n=158 narrow fields).

The country with the biggest Mendeley-Scopus outliers is Turkey (Figure 8). Both Rehabilitation (160 articles) and Orthopedics & Sports Medicine (312 articles) have much higher male citation impact but much lower male readership impact. For Rehabilitation, the journals in the category point to gender specialism differences within this field. Whilst most journals have mainly or exclusively male authors (e.g., European Journal of Orthopaedic Surgery and Traumatology: 1 female, 23 males), the biggest exceptions are Journal of Back and Musculoskeletal Rehabilitation (7 females, 8 males) and Fizyoterapi Rehabilitasyon (12 females, 1 male). This suggests that the rehabilitation-related research in this category may be more Mendeley-friendly, whereas the more surgery-related topics in this category are less Mendeley-friendly. Practicing doctors may avoid reference managers or use an alternative. The same explanation holds for Orthopedics & Sports Medicine. For example, Turkiye Fiziksel Tip ve Rehabilitasyon Dergisi (physical medicine and rehabilitation: 32 females, 20 males) is female and Eklem Hastaliklari ve Cerrahisi (joint diseases and surgery: 1 female, 22 males) is male and has a medical focus.

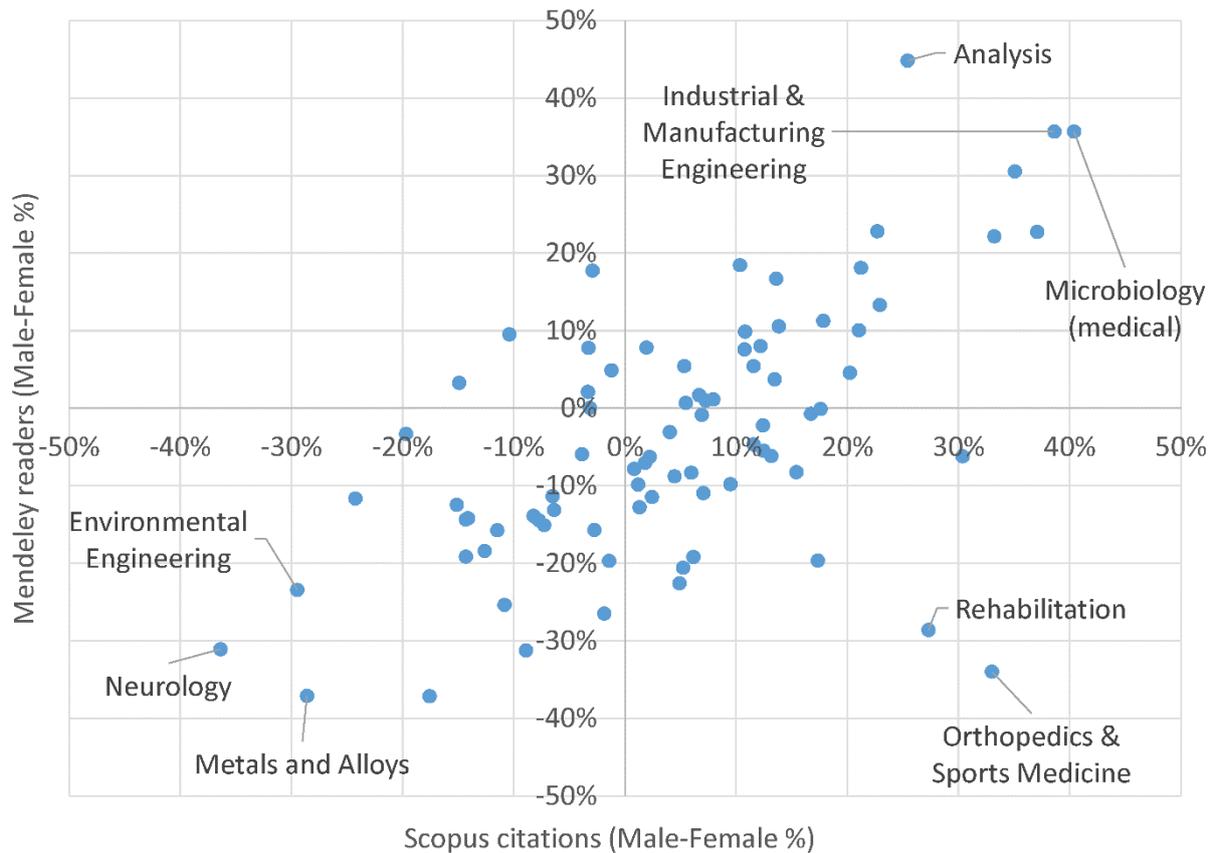

Figure 8. The average number of Mendeley readers per article difference between male-first-authored and female-first-authored Turkish articles from 2014 by narrow field (qualification: at least 50 male and 50 female articles in a field; n=82 narrow fields).

## 4 Discussion

This study is limited by being restricted to only five countries and one year. Substantially different results could be expected for other countries with differing gender expectations and differing levels of uptake of Mendeley. The data is likely to disproportionately exclude international researchers and so differences found between countries may be overestimated. In contrast, the method of attributing the gender of a paper solely to its first author may have a damping effect on the results for each individual country because some of the papers are likely to have substantial contributions from authors with genders that are different to that of the first author. Thus, assuming a simple model that a paper's citations are proportional to the contributions of genders to it, the method here is likely to underestimate the underlying gender difference within each country. There may also be characteristics of mixed gender author papers that are not the sum of the characteristics of the genders involved, however.

The citation results conflict with some prior studies that found a male citation advantage. The most likely reason for the difference is the use of arithmetic mean-based normalisation in all prior studies, which gives a substantial male advantage in comparison to geometric mean-based normalisation (discussed further below). Another important factor is the national scope of the analyses. Since the results here differed greatly between nations, findings for one country cannot be generalised to all others and international findings do not necessarily apply to individual countries. The gender differences might also be due to more recent data analysed here in some cases. They seem unlikely to be a result of the use here of the narrowest field classifications yet because the broader classification results advantaged females. The Scopus classification scheme is a possible reason for the differences since previous studies have used curated collections of journals, WoS categories or NSF categories (derived from WoS). It is possible that either WoS or Scopus is more likely to mix micro-fields with differing impact and author gender characteristics. It is also possible that the greater journal coverage of Scopus is a factor. Nevertheless, the extra Scopus journals would presumably have a lower average impact, and may attract a higher share of female authors, suggesting that Scopus would bias the results in favour of males, compared to WoS (by including additional low impact female authored articles).

The higher citation advantage of males using the MNCS compared to the MNLCS is an important new result. Using the arithmetic mean to average citation data has been argued against based on statistical inappropriateness because geometric means or log-based equivalents (e.g., Lundberg, 2007) are more precise for skewed data sets (Thelwall, 2016; Zitt, 2012). The increased precision is due to the ability of individual very high values to dramatically change the overall result. This can be seen, for example, in cases where a Journal Impact Factor jumps after publishing a single highly cited article. The issue in the results here is not precision, however, since the number of articles is large enough for the precision of the MNCS to be adequate, irrespective of skewing. Instead, the cause of the difference is that male-authored articles are more likely to have very high citation counts, consistent with a previous study of Norway (Aksnes, Rorstad, Piro, & Sivertsen, 2011). A non-statistical judgement is therefore necessary about whether it is better to allow highly cited papers to dominate the results. In favour of this, highly cited articles (excluding reviews) may make particularly influential contributions to research that their citations reflect. Conversely, highly cited papers may predominantly attract citations by imitation (de Solla Price, 1976; Merton, 1968), so that their citation counts overestimate their value. The reason why male authors produce a higher share of highly cited papers is unclear but it may be a side-effect of greater male career productivity (Aksnes, Rorstad, Piro, & Sivertsen, 2011).

There are many possible reasons for the female bias of Mendeley compared to Scopus (or the male bias of Scopus with respect to Mendeley). There is currently insufficient evidence to decide which, if any of the list below contribute.

- Mendeley may reflect a different type of impact to Scopus, such as with additional educational impact (from undergraduate users) or applied impact, with female-authored research tending to generate more of this type of impact.
- Higher male self-citation rates (King et al., 2017) may not translate into greater Mendeley self-readership rates.
- Almost equivalent to the above, males may be more likely to research in specialisms within narrow fields that tend to avoid Mendeley. This may include medical research (e.g., some doctors are reluctant to use the social web: Brown, Ryan, & Harris, 2014).
- Since prior gender biases against females have declined over time, younger researchers are more likely to be female than older researchers. Since Mendeley users probably tend to be younger than journal article authors, it may have a bias towards topics of interest for younger researchers, and hence disproportionately towards female-authored research.

The similar average citation impact of male-authored and female-authored research in Spain, the UK and the USA suggests that there is no substantial prejudice against female-authored articles in these countries, although it is also possible that there is some prejudice but it is counteracted by other factors that make female-authored research more impactful. The lower citation rates for female-authored research in India and Turkey could be due to many factors, including the following.

- Prejudice by male researchers against female researchers within India and Turkey (Gupta & Sharma, 2003; Gupta, 2017; Küskü, Özbilgin, & Özkale, 2007), combined with high levels of national self-citation in at least some research areas.
- Changing demographics in India and Turkey, with older (more male) researchers tending to produce higher impact research, or having more extensive networks of researchers that know their work and cite it.
- Female research topics tending to be in lower citation specialisms in India and Turkey, even within narrow fields.
- Female researchers having less access to resources in India and Turkey (e.g., funding), being less able to attend prestigious institutions (Gupta, 2012), or having additional socially-determined calls on their time (e.g., family responsibilities: Inanc & Özcan, 2016) or constraints on their working practices (e.g., being kept out of leadership roles [not Turkey], or being unable to access research networks due to perceived decorum issues (Gupta, 2007)).

## 5  Conclusions

The lack of a substantial citation bias to the detriment of females in Spain, the UK and the USA at the level of narrow fields gives some evidence that this issue should not be a priority in initiatives to overcome low numbers of females in some areas of science. Although prejudice against females is not necessarily the cause of the greater citation impact of male-authored research in India and Turkey, the results raise this possibility and provides evidence for people in these countries arguing that gender prejudice or informal bias (e.g., through cultural expectations) is an ongoing problem. The overall citation bias against females found in some previous studies may be due to mixing countries where there is little difference (UK, USA) with those with a more substantial difference, or to the use of at least partly older data. It is probably not due to aggregating at the level of broad fields since this tends to increase the relative impact of female first-authored research.

The citation advantage given to males by the MNCS calculations compared to the MNLCS show that the decision about which to use will have an important influence on the results of any gender study. This choice may be made on a theoretical basis with a decision about how to value highly cited research. If accepting high citation counts at face value, then MNCS is preferable, whereas if believing that they overestimate the underlying impact or quality of an article, then the MNLCS or geometric mean-based calculations are preferable. Given the historical legacy of prejudice against females in academia, it also seems reasonable to be swayed towards an indicator that better reflects the impact of their work.

The greater female impact in Mendeley compared to Scopus suggests that readers cannot be used as a proxy for citations (e.g., for early impact evidence: Thelwall, in press) when investigating gender differences. The lack of evidence about why the Mendeley readers give different gender inequality results to the Scopus citations raises the possibility that one may be superior in this regard. For example, Mendeley may reflect wider impacts (including educational uses) or narrower impacts (ignoring impact on subsequent research conducted by demographics that tend not to use Mendeley). The former seems more likely since similar results have been found for download data (Elsevier, 2017, p. 31), which is more universal than Scopus and Mendeley. Thus, the female-friendly results from Mendeley in four of the five countries examined raises the possibility that females tend to create *more* impactful research in most countries.